\documentclass[aps,pra,twocolumn,superscriptaddress,nobalancelastpage]{revtex4-1}

\usepackage{amsmath,amssymb,mathtools}
\usepackage[dvipdfmx]{graphicx}
\usepackage[colorlinks=true,bookmarks=true,citecolor=blue,linkcolor=blue,urlcolor=blue,breaklinks=true]{hyperref}

\graphicspath{{fig/}}
\usepackage{dcolumn}
\usepackage{bm}
\usepackage[T1]{fontenc}
\usepackage{textcomp}
\usepackage{color}
\usepackage{mathrsfs}
\usepackage[normalem]{ulem}

\begin{document}

\title{Magnonic Casimir Effect in Ferrimagnets}

\author{Kouki Nakata}
\email[]{{nakata.koki@jaea.go.jp}}
\thanks{equal contribution.}
\affiliation{Advanced Science Research Center, Japan Atomic Energy Agency, Tokai, Ibaraki 319-1195, Japan}

\author{Kei Suzuki}
\email[]{{k.suzuki.2010@th.phys.titech.ac.jp}}
\thanks{equal contribution.}
\affiliation{Advanced Science Research Center, Japan Atomic Energy Agency, Tokai, Ibaraki 319-1195, Japan}

\date{\today}

\begin{abstract}
Quantum fluctuations are the key concepts of quantum mechanics. Quantum fluctuations of quantum fields induce a zero-point energy shift under spatial boundary conditions. This quantum phenomenon, called the Casimir effect, has been attracting much attention beyond the hierarchy of energy scales, ranging from elementary particle physics to condensed matter physics together with photonics. However, the application of the Casimir effect to spintronics has not yet been investigated enough, particularly to ferrimagnetic thin films, although yttrium iron garnet (YIG) is one of the best platforms for spintronics. Here we fill this gap. Using the lattice field theory, we investigate the Casimir effect induced by quantum fields for magnons in insulating magnets and find that the magnonic Casimir effect can arise not only in antiferromagnets but also in ferrimagnets including YIG thin films. Our result suggests that YIG, the key ingredient of magnon-based spintronics, can serve also as a promising platform for manipulating and utilizing Casimir effects, called Casimir engineering. Microfabrication technology can control the thickness of thin films and realize the manipulation of the magnonic Casimir effect. Thus, we pave the way for magnonic Casimir engineering.
\end{abstract}

\maketitle


\textit{Introduction.}---Toward efficient transmission of information
that goes beyond what is offered by conventional electronics,
the last two decades have seen a significant development of 
magnon-based spintronics~\cite{MagnonSpintronics}, called magnonics.
The main aim of this research field is
to use the quantized spin waves, magnons,
as a carrier of information in units of the reduced Planck constant $\hbar$.
A promising strategy for this holy grail is to explore insulating magnets.
Thanks to the complete absence of any conducting metallic elements,
insulating magnets are free from drawbacks of conventional electronics, 
such as substantial energy loss due to Joule heating.
This is the advantage of insulating magnets.
Thus, exploring quantum functionalities of magnons in insulating magnets
is a central task in the field of magnonics.

\begin{figure}[t]
\centering
\includegraphics[width=0.47\textwidth]{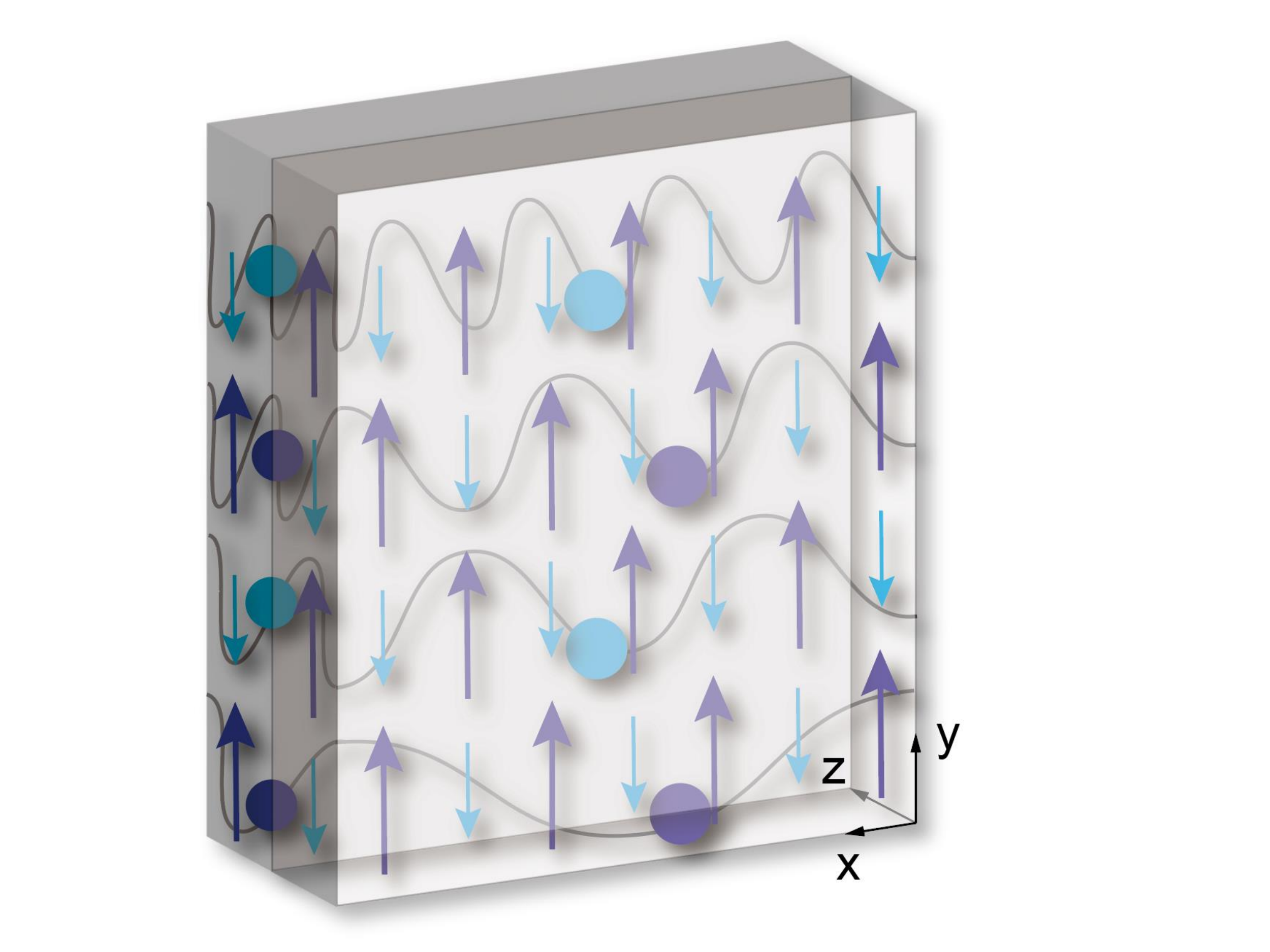}
\caption{Schematic of the ferrimagnetic thin film
for the magnonic Casimir effect.
Two kinds of magnons (circles) 
arise from the alternating structure of up and down spins (arrowed lines).
Wavy lines represent
quantum-mechanical behaviors of magnons in the discrete energy.}
\label{fig:YIGThinFilm}
\end{figure}

Quantum fluctuations of photon fields 
induce a zero-point energy shift,
called the Casimir energy, under spatial boundary conditions.
This Casimir effect is a fundamental phenomenon of quantum physics,
and the original platform for the Casimir effect was the photon field~\cite{CasimirEffect,CasimirExp,CasimirExpErra},
which is described by quantum electrodynamics.
The concept can be extended to various fields such as 
scalar, vector, tensor, and spinor fields.
Nowadays, the Casimir effect has been attracting much attention
beyond the hierarchy of energy scales, 
ranging from elementary particle physics
to condensed matter physics~\cite{CasimirNanoRMP,CasimirCosmoloReview}.
As an example, see Refs.~\cite{AFCasimir1989,CasimirAF1992,AFmagnonCasimir1998,SpinCasimir2015,RChengAFthermalCasimirMagnon,CasimirFM,SkyrmionFMCasimir,IvanovCasimirSoliton}
for Casimir effects in magnets~\footnote{See also Ref.~\cite{CasimirDynamicalSaito}
for an analog of the dynamical Casimir effect~\cite{DCE_Moore,DCE_Dodonov,DCE_Nori_RMP}
with magnon excitations in a spinor Bose-Einstein condensate.}.
However, the application of the Casimir effect 
to spintronics has not yet been studied enough,
particularly to ferrimagnetic thin films 
(see Fig.~\ref{fig:YIGThinFilm}),
although yttrium iron garnet (YIG)~\cite{YIGserga}
has been playing a central role in spintronics.

Here we fill this gap.
In terms of the lattice field theory,
we investigate the Casimir effect 
induced by quantum fields for magnons in insulating magnets
and refer to it as the magnonic Casimir effect.
We study the behavior of the magnonic Casimir effect
with a particular focus on the thickness dependence of thin films,
which can be experimentally controlled 
by microfabrication technology~\cite{BisaikakouYIG,BisaikakouYIG2}.
Then, we show that the magnonic Casimir effect can arise
not only in antiferromagnets (AFMs) but also in ferrimagnets 
with realistic model parameters for YIG thin films.
Our study indicates that YIG, an ideal platform for magnonics,
can serve also as a key ingredient of 
Casimir engineering~\cite{Review_CasimirEngineering}
which aims at exploring quantum-mechanical functionalities of nanoscale devices.


\textit{Antiferromagnets.}---We consider insulating AFMs 
described by the quantum Heisenberg Hamiltonian 
which has U$(1)$ symmetry about the quantization axis
and study the behavior of the magnonic Casimir effect
with a focus on the thickness dependence.
The AFM is a two-sublattice system,
and the ground state has the N\'eel magnetic order~\footnote{
As an example, Ref.~\cite{huang2017edge} studied magnon dynamics 
at the zigzag edge of a honeycomb lattice 
with long-ranged N\'eel magnetic order.}.
From the spin-wave theory with the Bogoliubov transformation, 
elementary excitations are two kinds of magnons 
designated by the index $\sigma=\pm$ 
having the spin angular momentum $\sigma \hbar$.
Owing to the U$(1)$ symmetry, 
the Hamiltonian can be recast into the diagonal form 
with the magnon energy dispersion for the wave number 
${\mathbf{k}}=(k_x, k_y, k_z)  $ as
$\epsilon_{\sigma,{\mathbf{k}}}$
where the total spin angular momentum of magnons is conserved.
Two kinds of magnons ($\sigma=\pm$) are in degenerate states.
Hence, we study the low-energy magnon dynamics of the insulating AFM 
by using the quantum field theory of complex scalar fields, 
i.e., the complex Klein-Gordon field theory~\cite{peskin,Ezawa}.
Then, we can see that there exists 
a zero-point energy~\cite{AndersonAF}.
This is the origin of the Casimir effect.
Note that throughout this study, 
we focus on clean insulating magnets and work under the assumption that
the total spin along the quantization direction is conserved 
and thus remains a good quantum number.

Through the lattice regularization,
the Casimir energy $E_{\text{Cas}} $
is defined as the difference between the zero-point energy
$ E_0^{\text{int}}$ 
for the continuous energy $ \epsilon_{\sigma,{\mathbf{k}}} $
and the one $ E_0^{\text{sum}} $ 
for the discrete energy $  \epsilon_{\sigma,{\mathbf{k}},n}$
with $ n \in {\mathbb{Z}}$.
In two-sublattice systems,
such as AFMs and ferrimagnets 
(see Fig.~\ref{fig:YIGThinFilm}),
the wave numbers on the lattice are replaced by
$ (k_j {a})^2 
\rightarrow 
2[1-\text{cos}(k_j {a})]$
in the $j$ direction for $j=x,y,z$,
where 
${a}$ is the length of a magnetic unit cell.
Here, by taking the Brillouin zone (BZ) into account,
we set the boundary condition for the $z$ axis in wave number space
$ {\mathbf{k}}=(k_x, k_y, k_z) $ as
$k_z \rightarrow \pi n/{L_z} $,
i.e., $ k_z {a} \rightarrow \pi n/{N_z} $,
where 
$L_z := {a} N_z $ is the thickness of magnets,
$ N_{j} \in {\mathbb{N}}$ is the number of magnetic unit cells 
in the $j$ direction for $j=x,y,z$,
and $n=1,2,..., 2N$ for $ N \in {\mathbb{N}}$.
The number of unit cells on the $xy$ plane is $ 4 N_x N_y$,
and that of magnetic unit cells is $ N_x N_y$.
Thus, the magnonic Casimir energy 
per the number of magnetic unit cells $ N_x N_y$ on the surface 
for $N_z=N$ is described as~\cite{actor2000casimir,pawellek2013finite,CasimirKS3,CasimirKS4,KSremnantCasimir}
\begin{subequations}
\begin{align}
 E_{\text{Cas}} 
 &:= E_0^{\text{sum}}(N)- E_0^{\text{int}}(N),  
\label{eqn:CasE} \\
 E_0^{\text{sum}}(N)
&=\sum_{\sigma=\pm}
\int_{\text{BZ}}
\frac{d^2 (k_{\perp} {a})}{(2\pi)^2}
\Bigg[
\frac{1}{2}
\Big(
\frac{1}{2}
\sum_{n=1}^{2N}
\lvert 
\epsilon_{\sigma,{\mathbf{k}},n}
\rvert 
\Big)
\Bigg], 
\label{eqn:CasEdisc} \\
 E_0^{\text{int}}(N)
&=
\sum_{\sigma=\pm}
\int_{\text{BZ}}
\frac{d^2 (k_{\perp} {a})}{(2\pi)^2}
\Bigg[
\frac{1}{2}
N
\int_{\text{BZ}}
\frac{d (k_z {a})}{2\pi}
\lvert 
\epsilon_{\sigma,{\mathbf{k}}}
\rvert 
\Bigg],
  \label{eqn:CasEcont}
\end{align}
\end{subequations}
where
$ k_{\perp}:=\sqrt{k_x^2+k_y^2} $,
$ d^2 (k_{\perp} {a})=d(k_x {a}) d(k_y {a}) $,
the integral is over the first BZ, and 
the factor $1/2$ arises from the zero-point energy for the scalar field.
Since the constant terms which are independent of the wave number
do not contribute to the Casimir energy,
we drop them throughout this study.
To see the dependence of the Casimir energy $E_{\text{Cas}} $
on the thickness of magnets $L_z := {a} N_z $,
it is convenient to introduce the rescaled Casimir energy 
$ C_{{\text{Cas}}}^{[b]} $
in terms of $N_z^{b}  $ for $ b \in {\mathbb{R}}$ as
\begin{align}
C_{{\text{Cas}}}^{[b]}:= 
 E_{\text{Cas}} \times  N_z^{b}.
  \label{eqn:CasCoefficient}
\end{align}
Then, we refer to $ C_{{\text{Cas}}}^{[b]} $
as the magnonic Casimir coefficient in the sense that
$E_{\text{Cas}} = C_{{\text{Cas}}}^{[b]} N_z^{-b}$.

\begin{figure}[t]
\centering
\includegraphics[width=0.49\textwidth]{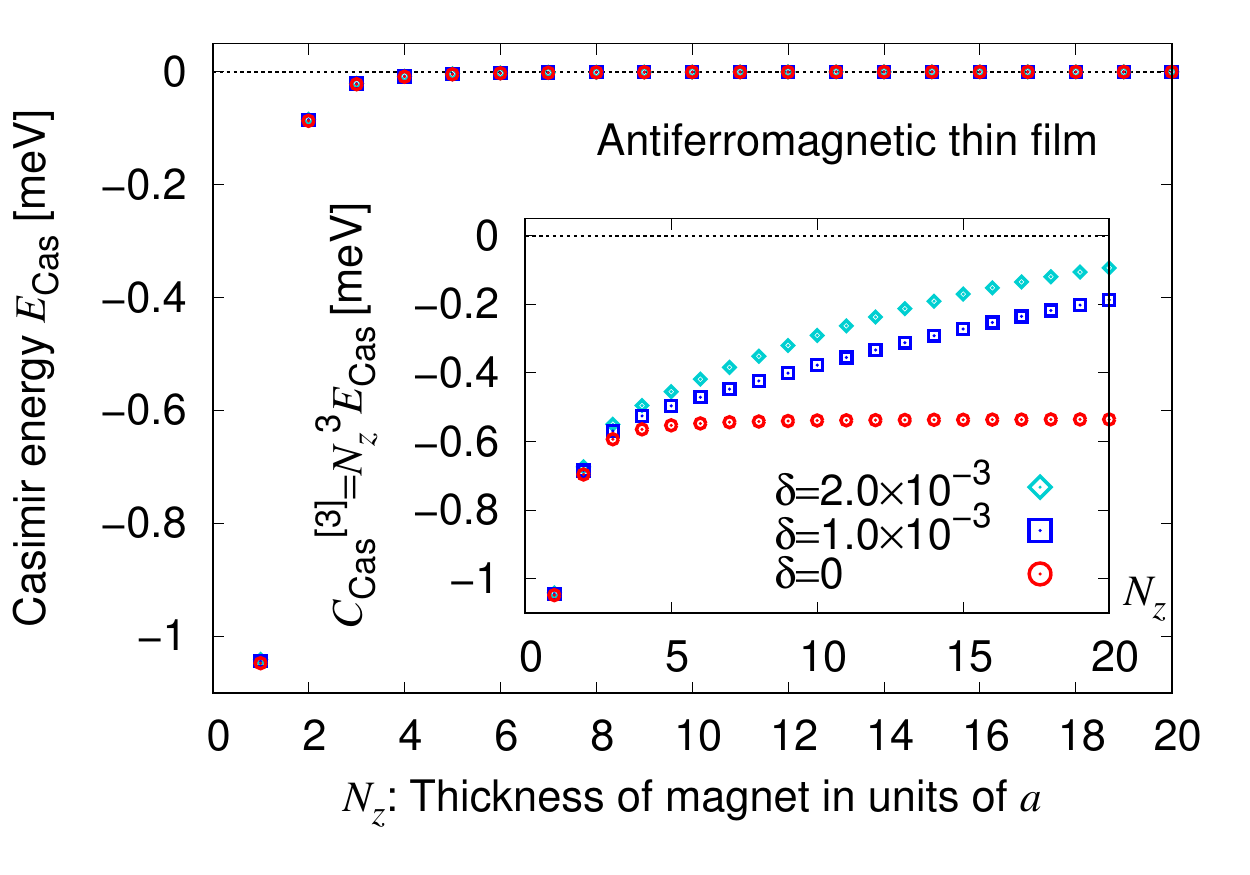}
\caption{Plots of the magnonic Casimir energy 
$ E_{\text{Cas}} $
and its coefficient (inset)
$C_{{\text{Cas}}}^{[3]} = E_{\text{Cas}} \times  N_z^{3}$
in the AFM [see Eq.~\eqref{eqn:EAF}]
as a function of $N_z$
for the thickness of magnets
$L_z={a} N_z$.}
\label{fig:AF}
\end{figure}

Here, we consider magnons in AFMs on a cubic lattice 
with the energy dispersion 
$\epsilon_{\sigma,{\mathbf{k}}}
= \epsilon_{\sigma,{\mathbf{k}}}^{\text{AFM}} $~\cite{KSJD}:
\begin{subequations}
\begin{align}
\epsilon_{\sigma,{\mathbf{k}}}^{\text{AFM}}
=& \hbar \omega_{0}
 \sqrt{\delta
 +\Big(\frac{k {a}}{2}\Big)^2},
 \label{eqn:EAF}  \\
\hbar \omega_{0}
:=&2\sqrt{3}JS,  
\label{eqn:omega0}   \\
\delta
:=&
3\Big[\Big(\frac{K}{6J}\Big)^2+2\Big(\frac{K}{6J}\Big) \Big],
\label{eqn:Delta}
\end{align}
\end{subequations}
where 
$ k:= | {\mathbf{k}} | $,
$J>0$ parametrizes the exchange interaction 
between the nearest-neighbor spins of the spin quantum number $S$,
and $K\geq 0$ is the easy-axis anisotropy,
which provides the magnon energy gap and ensures the N\'eel magnetic order.
Two kinds of magnons ($\sigma=\pm$) are in degenerate states.
In the absence of the spin anisotropy, $K=0$,
the energy gap vanishes $\delta=0$,
and the gapless magnon mode behaves like
a relativistic particle with the linear energy dispersion.
From the results obtained in Refs.~\cite{CrOexp,CrOexp2,Cr2O3_LatticeConstant},
we roughly estimate 
the model parameter values for ${\text{Cr}}_2{\text{O}}_3$,
as an example, as follows~\cite{KSJD}:
$J=15$ meV,
$S=3/2$,
$K=0.03$ meV,
and
$a=0.496\,07$ nm.
These parameter values provide
$ \hbar \omega_{0} \sim  77.94$ meV
and
$\delta   \sim  2 \times 10^{-3} $.

Figure~\ref{fig:AF} shows that 
the magnonic Casimir effect arises in the thin film of the AFM.
The magnonic Casimir energy 
$ E_{\text{Cas}} $
of the magnitude {$O(10^{-2})$} meV 
is induced for $N_z \geq 2$.
Even in the presence of the magnon energy gap $\delta \neq 0$,
the absolute value amounts to 
{$O(10^{-2})$} meV
and decreases as the magnon energy gap increases.
Thus, the magnonic Casimir energy takes a maximum absolute value 
in the gapless mode $\delta=0$, 
where the magnon behaves like a relativistic particle 
with the linear energy dispersion.
We remark that in the case of the gapped magnon modes,
the absolute value of the magnonic Casimir coefficient 
$  C_{{\text{Cas}}}^{[3]} = E_{\text{Cas}} \times  N_z^{3} $
decreases as the thickness of the thin film increases.
This behavior is similar to the Casimir effect known for 
massive degrees of freedom~\cite{hays1979vacuum,ambjorn1983properties}.
In the case of the gapless mode,
the magnonic Casimir coefficient
$  C_{{\text{Cas}}}^{[3]} $
approaches asymptotically to a constant value,
and the magnonic Casimir energy exhibits the behavior of
$  E_{\text{Cas}} \propto  1/N_z^{3} $
as the thickness increases.
The asymptotic value of 
$C_{{\text{Cas}}}^{[3]}$
for the gapless magnon mode $\delta=0$ 
given in the numerical result 
(see Fig.~\ref{fig:AF})
is estimated approximately as 
{$(-\pi^2/720) \times (\hbar \omega_{0}/2) 
\sim -0.5341$ meV}
from an analytical calculation.
The factor of {$-\pi^2/720$} is well known as the analytic solution 
for the conventional Casimir effect of a massless complex scalar field 
in continuous space~\cite{ambjorn1983properties}.
Thus, although the magnonic Casimir effect is realized on the lattice, 
it is qualitatively and quantitatively analogous to 
the continuous counterpart, 
except for $a$-dependent lattice effects.


\textit{Ferrimagnets.}---We develop the study of AFMs into ferrimagnets
where the ground state has an alternating structure 
of up and down spins on a cubic lattice 
(see Fig.~\ref{fig:YIGThinFilm}).
In contrast to AFMs, 
the spin quantum number on the two-sublattice 
is different from each other in ferrimagnets.
Hence, the degeneracy for two kinds of magnons ($\sigma=\pm$)
is intrinsically lifted.
In ferrimagnetic thin films, 
dipolar interactions due to the nonzero magnetization 
play a key role.
Still, at low temperatures 
where the magnon-magnon interaction of the fourth order 
in magnon operators is negligibly small~\cite{YIGthin2008},
the number of magnons and the total spin angular momentum are conserved,
and the Hamiltonian for the ferrimagnetic thin film
can be diagonalized with the magnon energy dispersion
$\epsilon_{\sigma,{\mathbf{k}}}
= \epsilon_{\sigma,{\mathbf{k}}}^{\text{ferri}}$.

Here, we consider magnons 
in the thin film of clean insulating ferrimagnets on a cubic lattice 
subjected to in-plane magnetic fields at such low temperatures.
Still, due to the competition between dipolar and exchange interactions,
the minimum energy point shifts from the zero mode of magnons,
$  {\mathbf{k}}=0 $,
to a finite wave number mode 
which is characterized by the thickness of the thin film
$L_z = {a} N_z $ (see Fig.~\ref{fig:YIGThinFilm}).

\begin{figure}[t]
\centering
\includegraphics[width=0.49\textwidth]{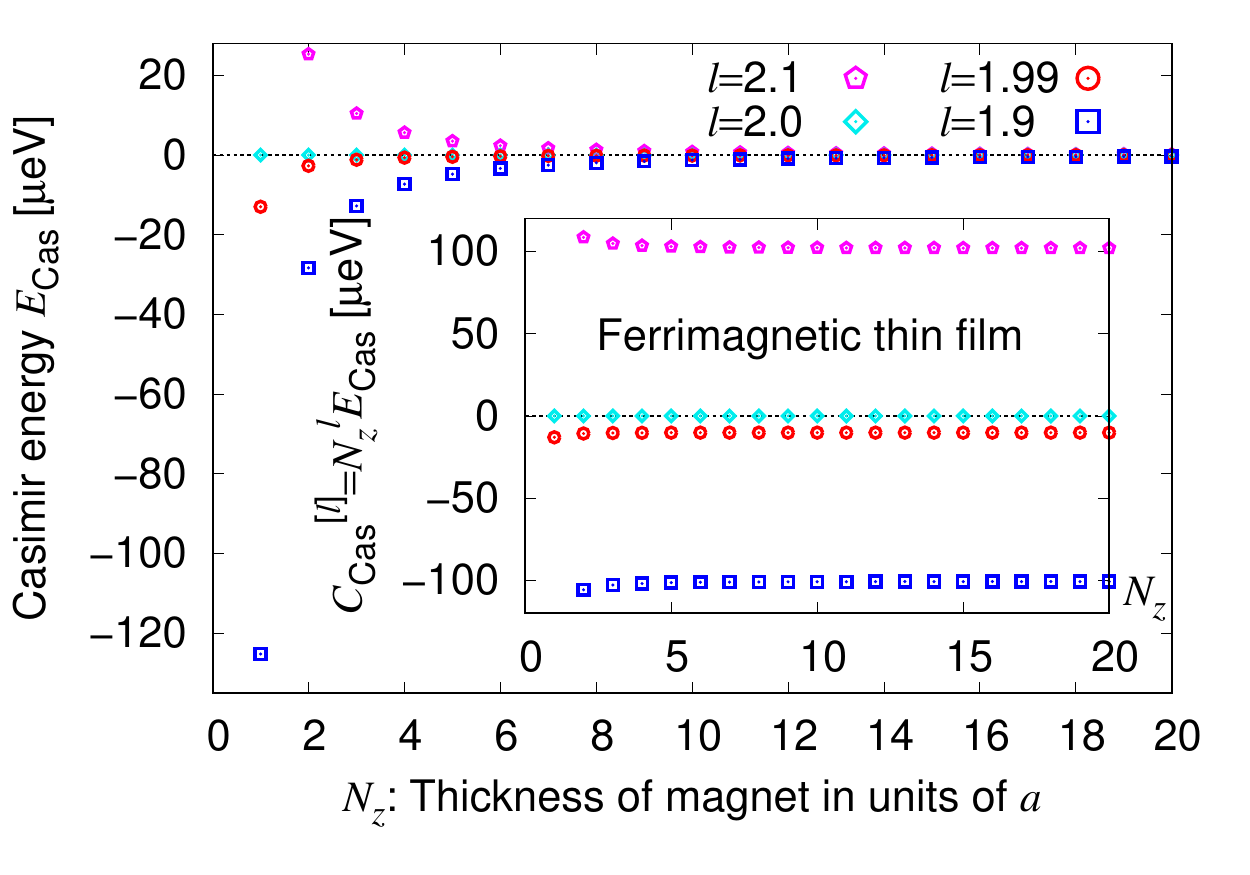}
\caption{Plots of the magnonic Casimir energy
$E_{\text{Cas}} $
and its coefficient (inset)
$  C_{{\text{Cas}}}^{[l]} = E_{\text{Cas}} \times  N_z^{l} $
for $l=2.1$,
$l=2.0$,
$l=1.99$,
and $l=1.9$
in the ferrimagnetic thin film [see Eq.~\eqref{eqn:Eferri}]
as a function of $N_z$
for the thickness of magnets $L_z={a} N_z$
with the model parameter values for YIG of $D_z=D$.}
\label{fig:YIG}
\end{figure}

The magnon energy dispersion along the in-plane direction
is provided in Refs.~\cite{SlavinYIG,SlavinYIG2},
whereas the dispersion along the $z$ axis in the thin film
has not yet been established~\cite{RezendeTextbook}.
Hence, taking into account the competition 
between dipolar and exchange interactions in the thin film,
we phenomenologically assume the behavior that 
the power of $k_z$, $ l \in {\mathbb{R}} $,
approaches asymptotically to $l=2$ in the bulk limit,
whereas it slightly differs from $l=2$
as long as we consider the thin film 
(see Fig.~\ref{fig:YIGThinFilm}).
Using this assumption and Refs.~\cite{SlavinYIG,SlavinYIG2},
the magnon energy dispersions are
\begin{subequations}
\begin{align}
\epsilon_{\sigma,{\mathbf{k}}}^{\text{ferri}}
=&\sqrt{\sigma H_0+\Delta_{\sigma}
{+\frac{D}{{a}^2} (k_{\perp} {a})^2
+\frac{D_z}{{a}^2}
(|k_z| {a})^l}}  
\label{eqn:Eferri}  \\
\times &
\sqrt{\sigma H_0+\Delta_{\sigma}
{+\frac{D}{{a}^2} (k_{\perp} {a})^2
+\frac{D_z}{{a}^2} (|k_z| {a})^l}
+ \sigma \hbar \omega_{{M}} {\mathcal{F}}_{\mathbf{k}}},
\nonumber \\
{\mathcal{F}}_{\mathbf{k}}
:=&
{\mathcal{P}}_{ k_{\perp}}(1-{\mathcal{P}}_{ k_{\perp}})
\frac{\sigma \hbar \omega_{{M}}}{\sigma H_0+\Delta_{\sigma}+\frac{D}{{a}^2} (k_{\perp} {a})^2 +\frac{D_z}{{a}^2} (|k_z| {a})^l}
\Big(\frac{{k_x}}{k_{\perp}}\Big)^2   
\nonumber    \\
+&1-{\mathcal{P}}_{ k_{\perp}}
\Big(\frac{{k_y}}{k_{\perp}}\Big)^2,
\label{eqn:F} \\
{{\mathcal{P}}_{ k_{\perp}}}
:=&
1-
\frac{1-{\text{e}}^{-k_{\perp} L_z}}{k_{\perp} L_z},
\label{eqn:P}
\end{align}
\end{subequations}
where the external magnetic field is applied 
along the $y$ axis
(see Fig.~\ref{fig:YIGThinFilm})
and $ \sigma H_0$ represents the resulting Zeeman energy,
$\Delta_{\sigma} \geq 0$
is the (intrinsic) magnon energy gap in ferrimagnets,
$D_{(z)}>0$ is the spin stiffness constant,
$ k_{\perp}:=\sqrt{k_x^2+k_y^2}  $,
{$\hbar \omega_{{M}}:=4\pi \gamma M_{{s}}$}
with the saturated magnetization density $M_{{s}} $
and the gyromagnetic ratio $\gamma $,
and the term
$  {\mathcal{F}}_{\mathbf{k}}  $
is responsible for the shift of the minimum energy point
from the zero mode to a finite wave number mode 
due to the competition between dipolar and exchange interactions
in the ferrimagnetic thin film: 
The first term of $ {\mathcal{F}}_{\mathbf{k}}  $ 
[see Eq.~\eqref{eqn:F}]
reproduces the Damon-Eshbach magnetostatic surface mode~\cite{DamonEshbach},
and the last term reproduces the backward volume magnetostatic mode~\cite{DamonEshbach}.

From the results obtained in
Refs.~\cite{YIGexpValue,YIGexpValue2,YIGexpValue3},
the model parameter values for YIG thin films 
are estimated as follows:
$D/{a}^2 \sim 3.376\,45 $ meV~\cite{YIGthin2008}
with $a=1.2376$ nm~\cite{YIGlatticeconstant1957},
$H_0 \sim8.103\,73 \  \mu$eV~\cite{YIGthin2008}, and
$\hbar \omega_{{M}} \sim20.3369 \   \mu$eV~\cite{YIGthin2008}.
Then, we estimate the magnon energy gap $\Delta_{\sigma} $ 
as~\footnote{The magnon energy dispersions and their temperature dependence in YIG were measured by inelastic neutron scattering~\cite{YIGdata,YIGdata2,ShamotoPRB,NambuYIG,ShamotoYIG,YIGdata3,YIGspectrum2021}.
We estimate the magnon energy gap $\Delta_{\sigma} $
by applying the model~\cite{KNST_OBarnett} of the effective block spins to YIG~\cite{YIGthin2008,ShamotoYIG}.
The theoretical estimate for the value,
$\Delta_{\sigma=-} - \Delta_{\sigma=+} \sim 39.848\,81 $ meV,
is consistent with the experimental data~\cite{NambuYIG}.}
$\Delta_{\sigma=-} - \Delta_{\sigma=+} \sim 39.848\,81 $ meV
with
$ \Delta_{\sigma=+} \sim 2.131\,91  $ meV 
and $ \Delta_{\sigma=-} \sim   41.980\,72$ meV,
which satisfy the condition
$  \Delta_{\sigma} \gg   \hbar \omega_{{M}} $.
In this condition, 
we study the low-energy magnon dynamics 
of the ferrimagnetic thin film
by using the quantum field theory of real scalar fields, 
i.e., the real Klein-Gordon field theory~\cite{peskin,Ezawa}.
Then, we can see that there exists a zero-point energy~\cite{YIGthin2008}.
The magnonic Casimir energy through the lattice regularization 
is given as Eq.~\eqref{eqn:CasE}.
We remark that the zero-point energy arises from quantum fluctuations
and does exist even at zero temperature.
The zero-point energy defined at zero temperature
does not depend on the Bose-distribution function
[Eqs.~\eqref{eqn:CasEdisc} and~\eqref{eqn:CasEcont}].
Hence, not only the low-energy mode ($\sigma=+$)
but also the high-energy mode ($\sigma=-$)
contribute~\footnote{Higher energy bands than 
those of Eq.~\eqref{eqn:Eferri}
also contribute to the magnonic Casimir energy.
However, the contribution becomes smaller
as the shape of the bands is flatter.
Numerical calculations of Refs.~\cite{YIGdata3,ShamotoYIG,YIGspectrum2021} show that 
higher energy bands tend to be flat.
Thus, we expect that the magnonic Casimir energy is dominated by
the two bands of Eq.~\eqref{eqn:Eferri}.} 
to the magnonic Casimir energy.

Under the phenomenological assumption that 
the value of $l$ [see Eq.~\eqref{eqn:Eferri}]
approaches asymptotically to $l=2$ in the bulk limit,
in this work focusing on the thin film,
we study the behavior of the magnonic Casimir effect
by changing the value slightly from $l=2$.
As examples, we consider the cases of 
$l=2.1$,
$l=1.99$,
and $l=1.9$.
Figure~\ref{fig:YIG} shows that the magnonic Casimir effect arises
in the ferrimagnetic thin film.
There is the magnonic Casimir energy $E_{\text{Cas}}$ of the magnitude 
{$O(1)$,
$O(10)$,
and $O(10) \ \mu$eV}
for
$l=1.99$, 
$l=1.9$, 
and $l=2.1$,
respectively, 
in $N_z \geq 2$.
As the thickness increases,
the magnonic Casimir coefficient $  C_{{\text{Cas}}}^{[l]} $
approaches asymptotically to a constant value,
and the magnonic Casimir energy exhibits the behavior of
$  E_{\text{Cas}} \propto  1/N_z^{l} $.
We also find from Fig.~\ref{fig:YIG} that 
the sign of the magnonic Casimir coefficient 
and energy for $l=2.1$ is positive
$C_{{\text{Cas}}}^{[2.1]}= E_{\text{Cas}} \times  N_z^{2.1} >0$
in $N_z \geq 2$,
whereas that for $l=1.9$ is negative
$C_{{\text{Cas}}}^{[1.9]}= E_{\text{Cas}} \times  N_z^{1.9} <0$.
This means that the Casimir force works in the opposite direction.

Note that even if $l=2$,
the magnonic Casimir effect arises in the ferrimagnetic thin film.
We have numerically confirmed that although the value is small,
there does exist the magnonic Casimir energy of the magnitude
{$ | E_{\text{Cas}} | \leq O(0.1)$ neV}
for $l=2$.
This strong suppression of the Casimir energy 
is a general property of the Casimir effect 
for quadratic dispersions on the lattice~\cite{KSremnantCasimir},
and the survival values originate from 
the dipolar interaction in the ferrimagnetic thin film.

We remark that as long as we consider thin films,
the value of $D_z$ can differ from $D$.
Even in that case, the magnonic Casimir effect arises.
When the value of $D_z$ changes from $D$ to $0.8 D$ as an example,
the magnonic Casimir energy $E_{\text{Cas}}$ 
increases approximately by $0.8$ times.
For more details about its dependence on 
the parameters $D_z$ and $l$,
see the Supplemental Material.


\textit{Proposal for experimental observation.}---The magnonic Casimir energy of
the ferrimagnetic thin film
depends strongly on
external magnetic fields through Zeeman coupling 
as in Eq.~\eqref{eqn:Eferri} 
and contributes to magnetization of magnets,
whereas the photon and phonon Casimir effects~\footnote{See Refs.~\cite{PCE_2014_Kamenev,PCE_2018_Efremov,PCE_2019_Efremov,PCE_2019_Rodin,PCE_2021_Rodin} for the Casimir effect of phonons and, e.g., Ref~\cite{DCE_Nori_PRA} for the dynamical one.} do not usually.
On the other hand, 
in the presence of magnetostriction~\cite{smith1963magnetostriction,PhysRev.130.1735,dudko1971magnetostriction,yacovitch1977magnetostriction}, 
the phonon Casimir effect is influenced by magnetostriction,
and its correction for the phonon Casimir energy depends on magnetic fields and contributes to magnetization.
However, such a contribution to magnetization 
from the phonon Casimir effect 
should be negligibly small by the factor of $10^{-6}$ 
compared with that from the magnonic Casimir effect
of ferrimagnets
because the magnetostriction constant
(i.e., the correction for the lattice constant)
is known to be $10^{-6}$ for
YIG~\cite{smith1963magnetostriction,PhysRev.130.1735}.
Hence, even in the presence of magnetostriction, 
the magnonic Casimir effect can be distinguished 
from the others.
Thus, we expect that 
our theoretical prediction, 
the magnonic Casimir effect in ferrimagnets,
can be experimentally observed through measurement of magnetization
and its film thickness dependence.
For more details, see the Supplemental Material.

For observation,
a few comments are in order.
First, 
we remark on edge/surface magnon modes.
The magnonic Casimir effect in our setup 
(see the thin film of Fig.~\ref{fig:YIGThinFilm}) 
is induced by magnon fields 
with wave numbers $k_z$ discretized by small $N_z$,
and its necessary condition is a $k_z$-dependent dispersion relation
under the discretization of $k_z$.
Throughout this study, 
we consider thin films of $ N_z \ll N_x, N_y $.
Even if additional edge/surface magnon modes exist
as well as
the Damon-Eshbach magnetostatic surface mode
and 
the backward volume magnetostatic mode 
[see Eq.~\eqref{eqn:F}],
they are confined only on the $x$-$y$ plane.
Then, their wave number in the $z$ direction is always zero, 
i.e., $k_z=0$, 
and its energy dispersion relation is independent of $k_z$.
Since a $k_z$-independent dispersion relation cannot induce 
the Casimir effect, the edge/surface modes cannot contribute to 
the magnonic Casimir effect.
In this sense, our magnonic Casimir effect is not affected by 
the existence of edge/surface magnon modes.

Note that
details of the edge condition,
such as the presence or absence of disorder,
may change the boundary condition
for the wave function of magnons,
but the existence of the magnonic Casimir effect
remains unchanged.
Even if there is a change in the spectrum near the edge,
the magnonic Casimir effect is little influenced
as long as one does not assume an ultrathin film
such as $N_z = 1,2,3$.
In this sense,
we expect that the following size of thin films is appropriate for 
observation of our prediction:
$N_z \sim 10$,
i.e.,
the film thickness of YIG is
$L_z=a N_z \sim 12.376$ nm.
Note that 
microfabrication technology~\cite{BisaikakouYIG,BisaikakouYIG2}
can control the thickness of thin films
and realize the manipulation of the magnonic Casimir effect.

Next, 
we remark on the magnon band structure.
Since our magnonic Casimir effect is induced by 
the $k_z$-dependent dispersion,
its Casimir energy of ferrimagnets 
is mainly characterized by the $D_z$-term 
in Eq.~\eqref{eqn:Eferri},
i.e., $ D_z (|k_z|a)^l $.
Hence,
we have investigated its dependence on
both $l$ and $ D_z $
(see Fig.~\ref{fig:YIG} and the Supplemental Material).
Even if the magnon band structure is affected due to some reasons,
the magnonic Casimir effect of ferrimagnets
is little influenced
by other details of the magnon band structure
except for $l $ and $D_z $.

Lastly,
we remark on thermal effects.
At nonzero temperature, 
thermal contributions to the Helmholtz free energy arise.
However, at low temperatures compared to 
$\epsilon_{\sigma,{\mathbf{k}}}$~\footnote{Note that 
even at such low temperatures, 
magnons do not form Bose-Einstein condensates 
in equilibrium~\cite{bunkovBECarXiv}.},
the thermal contribution is exponentially suppressed
due to the Boltzmann factor and becomes negligibly small.
Hence, at such low temperatures,
the contribution of the magnonic Casimir energy 
given as Eq.~\eqref{eqn:CasE} is dominant.



\textit{Conclusion.}---We have shown that 
the magnonic Casimir effect can arise not only in antiferromagnets 
but also in ferrimagnets with realistic model parameters for YIG.
Since the lifetime of magnons in YIG thin films 
is the longest among known materials,
and magnons exhibit long-distance transport 
over centimeter distances~\cite{WeesNatPhys},
YIG is the key ingredient of magnonics~\cite{MagnonSpintronics,YIGserga},
which has already realized the magnon transistor~\cite{MagnonTransistor}.
Our result suggests that YIG can serve also as a promising platform 
for Casimir engineering~\cite{Review_CasimirEngineering}:
Because the magnonic Casimir effect contributes to 
the internal pressure of thin films,
it will provide the new principles of nanoscale devices 
such as highly sensitive pressure sensors and magnon transistors.
Thus, our study paves the way for magnonic Casimir engineering.


We thank 
Yasufumi Araki, 
Yoshinori Haga,
Masaki Imai, 
Se Kwon Kim,
and
Katsumasa Nakayama
for fruitful discussions.
We acknowledge support
by Leading Initiative for Excellent Young Researchers, 
MEXT, Japan (K.N.),
by JSPS KAKENHI 
Grants No. JP20K14420 (K. N.), 
No. JP22K03519 (K. N.),
No. JP17K14277 (K. S.), 
and No. JP20K14476 (K. S.).


\bibliography{PumpingRef}

\newpage

\begin{center}
\noindent{\large{\textbf{Supplemental Material}}}
\end{center}

In this Supplemental Material,
first,
we provide some details 
about the dependence of the magnonic Casimir effect
on the parameters $l$ and $D_z$
in ferrimagnetic thin films.
Next, we remark on its film thickness dependence.
Then, we provide another point of view
for its robustness against disorder effects.
Lastly, we comment on the distinction 
between the Casimir effect
and the thermal Casimir effect.

\section{The parameter $l$- and $D_z$-dependence}

In the main text,
we have studied the magnonic Casimir energy
$E_{\text{Cas}} $
and
the coefficient
$  C_{{\text{Cas}}}^{[l]} = E_{\text{Cas}} \times  N_z^{l} $
for 
$l=2.1$,
$l=2.0$,
$l=1.99$,
and $l=1.9$
in the ferrimagnetic thin film
by using the model parameter values for YIG 
with fixed $D_z=D$.
Here,
we provide more details
about its dependence on the parameters $l$ and $D_z$.

First, we consider the cases of
$l=1.5$
and
$l=1.0$
with setting $D_z=D$.
Figure~\ref{fig:YIG_sup_ldep} shows that 
the magnonic Casimir effect still arises
in the ferrimagnetic thin film.
There is the magnonic Casimir energy 
$ E_{\text{Cas}} $
of the magnitude 
{$O(10^{-2}) $ meV, 
$O(10^{-1}) $ meV,
and
$O(10^{-1}) $ meV}
for
$l=1.9$, 
$l=1.5$, 
and 
$l=1.0$,
respectively, 
in $  N_z \geq 2$.
As the value of $l$ decreases from $l=2$
and approaches to $l=1$,
the magnitude of the magnonic Casimir energy increases.
Note that it amounts to 
{$O(10^{-1})$ meV}
even in $N_z=O(10)$
for $l=1.0$.
As the thickness increases,
the magnonic Casimir coefficient
$  C_{{\text{Cas}}}^{[l]} $
approaches asymptotically to a constant value
and the magnonic Casimir energy
exhibits the behavior of
$  E_{\text{Cas}} \propto  1/N_z^l $.

Then,
we consider the cases of 
$D_z/D=0.3$,
$D_z/D=0.5$,
and $D_z/D=0.8$
by fixing
$l=1.99$.
Figure~\ref{fig:YIG_sup_Dzdep} shows that 
the magnonic Casimir effect still arises
in the ferrimagnetic thin film.
When the value of $D_z$ changes from
$D$ to $0.8 D$ as an example,
the magnonic Casimir energy $E_{\text{Cas}}$ 
increases approximately by $0.8$ times.
Thus, the value of $E_{\text{Cas}}$ 
is approximately proportional to $D_z$.

\renewcommand{\thefigure}{S\arabic{figure}}
\setcounter{figure}{0}
\begin{figure}[t]
\centering
\includegraphics[width=0.49\textwidth]{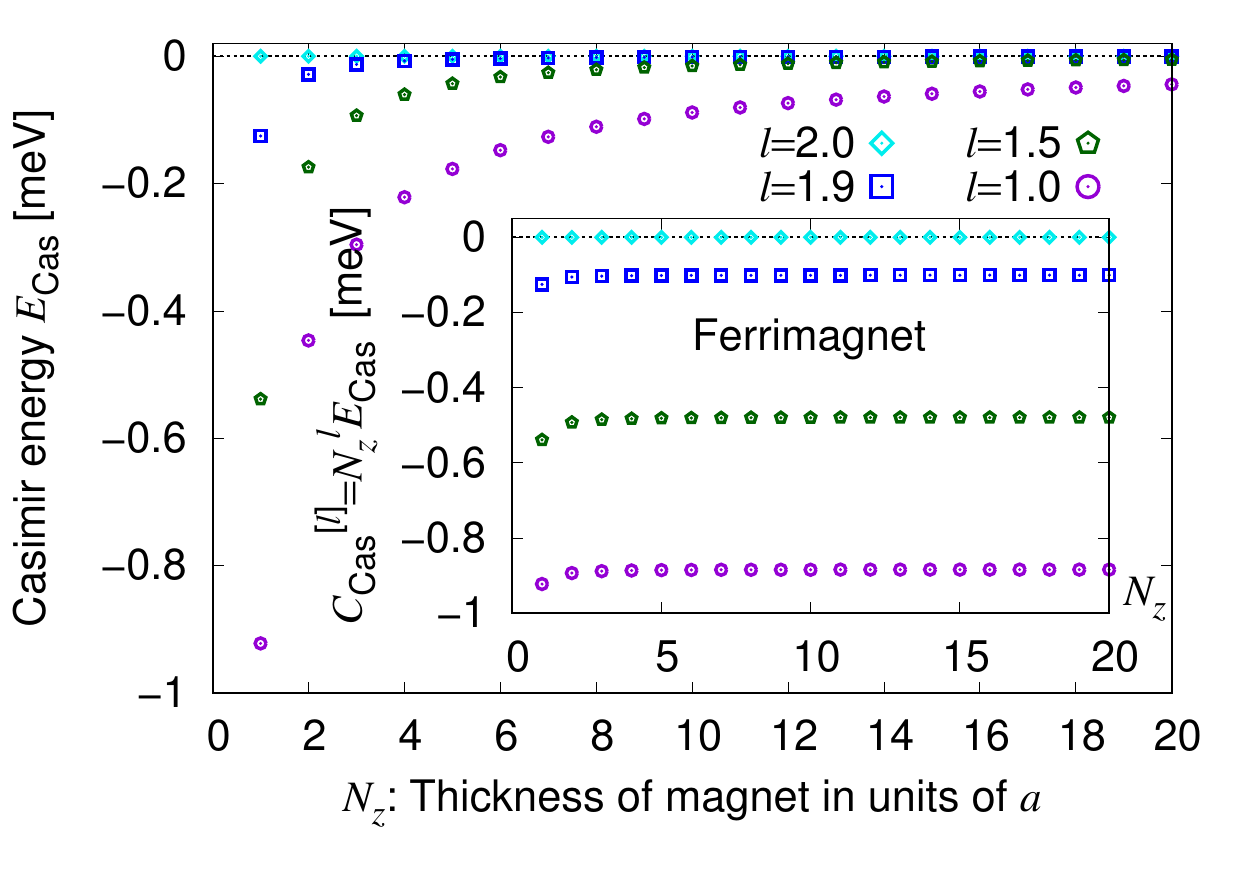}
\caption{Plots of the magnonic Casimir energy
$E_{\text{Cas}} $
and the coefficient (inset)
$  C_{{\text{Cas}}}^{[l]} = E_{\text{Cas}} \times  N_z^{l} $
for
$l=2.0$,
$l=1.9$,
$l=1.5$,
and $l=1.0$
in the ferrimagnetic thin film
as a function of $N_z$
for the thickness of magnets
$L_z={a} N_z$
under the model parameter values for YIG 
with fixed $D_z=D$.
}
\label{fig:YIG_sup_ldep}
\end{figure}

\begin{figure}[t]
\centering
\includegraphics[width=0.49\textwidth]{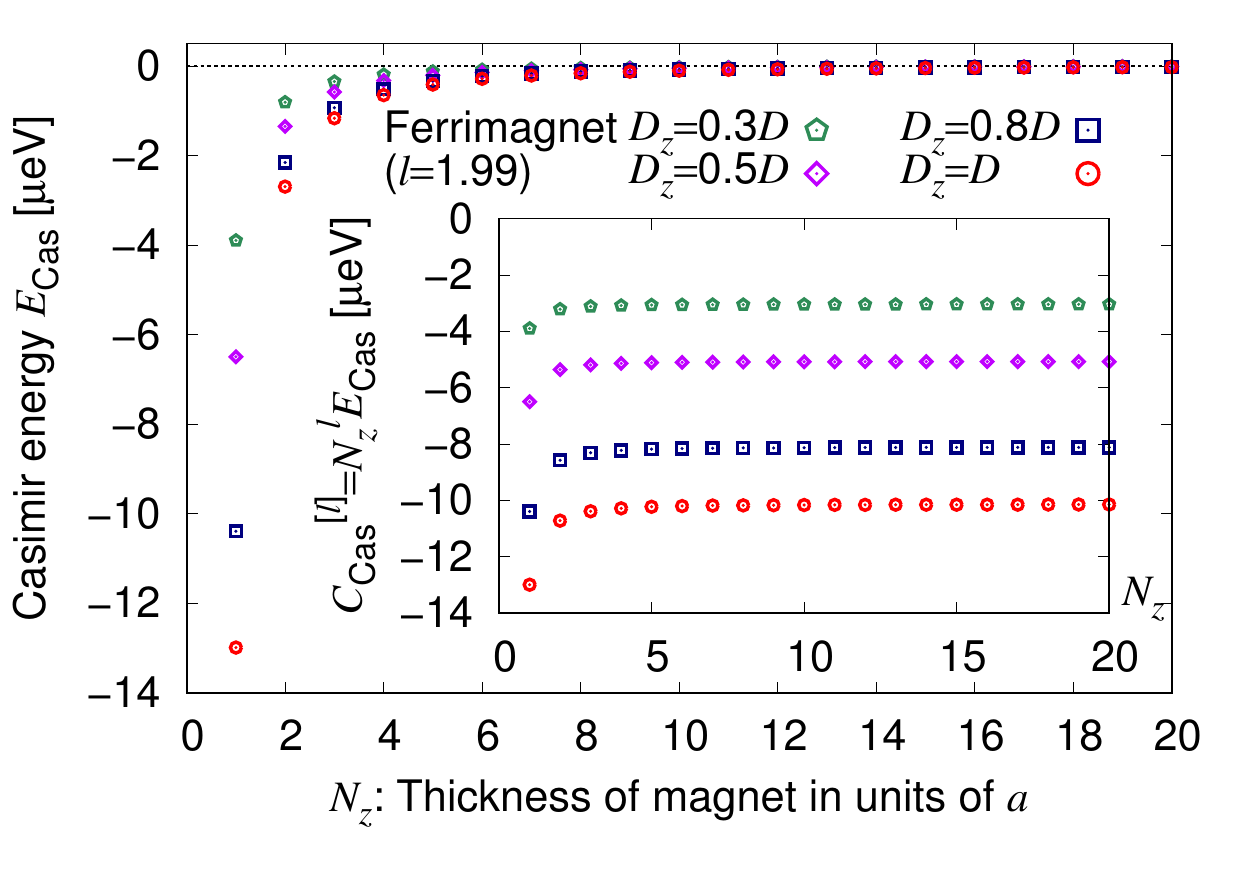}
\caption{Plots of the magnonic Casimir energy
$E_{\text{Cas}} $
and the coefficient (inset)
$  C_{{\text{Cas}}}^{[l]} = E_{\text{Cas}} \times  N_z^{l} $
for
$D_z/D=0.3$,
$D_z/D=0.5$,
$D_z/D=0.8$,
and $D_z/D=1.0$
in the ferrimagnetic thin film
as a function of $N_z$
for the thickness of magnets
$L_z={a} N_z$
under the model parameter values for YIG 
with $l=1.99$.
}
\label{fig:YIG_sup_Dzdep}
\end{figure}

\section{Remarks on the thickness dependence of magnetization}

In the main text,
we have remarked that our prediction,
the magnonic Casimir effect in ferrimagnets,
can be observed through measurement of magnetization
and its film thickness dependence.
Here, we add an explanation about it.
At zero temperature,
the Helmholtz free energy of magnon fields in thin films 
(i.e., the sum over discrete $ k_z$) 
is
$  E_0^{\text{sum}}(N_z) N_x N_y $,
and that under the bulk approximation 
(i.e., the integral with respect to continuous $k_z $) 
is
$ E_0^{\text{int}}(N_z) N_x N_y $ 
[see Eqs.~\eqref{eqn:CasE}-\eqref{eqn:CasEcont}].
The difference between them
is characterized by the magnonic Casimir energy 
$E_{\text{Cas}}$
as 
$  E_0^{\text{sum}}(N_z) N_x N_y
=  E_0^{\text{int}}(N_z) N_x N_y 
+  E_{\text{Cas}} N_x N_y $,
where the magnon energy dispersion of ferrimagnets 
(i.e., magnets including dipolar interactions) 
is Eq.~\eqref{eqn:Eferri}.
Note that the magnetic-field derivative 
(i.e., $H_0$-derivative)
of the Helmholtz free energy is magnetization.
Then, magnetization of thin films 
consists of two parts:
The bulk component and the magnonic Casimir energy.
Since 
$E_0^{\text{int}}(N_z) \propto  N_z $,
whereas
$  E_{\text{Cas}} \propto 1/(N_z)^l $,
magnetization of thin films exhibits 
a different $ N_z $-dependence from 
the bulk component,
and its difference is characterized 
by the magnonic Casimir energy.
In other words,
magnetization of thin films exhibits a different 
film thickness dependence from 
the bulk component
due to the magnonic Casimir effect.
Hence, our prediction, the magnonic Casimir effect 
in ferrimagnetic thin films 
(i.e., magnetic thin films including dipolar interactions),
can be observed through measurement of magnetization 
and its film thickness dependence.

Note that 
if dipolar interactions are relevant also in antiferromagnets,
its low-energy magnon dynamics is 
essentially described 
by Eq.~\eqref{eqn:Eferri} given for ferrimagnets.
The only difference is that 
the spin quantum number for each sublattice
is identical
in antiferromagnets,
where
the (intrinsic) magnon energy gap
for each mode $ \sigma=\pm $
can be identical
$ \Delta_{\sigma=+} = \Delta_{\sigma=-}$ 
[see Eq.~\eqref{eqn:Eferri}].
In this sense,
its Casimir effect
exhibits qualitatively the same behavior as Fig.~\ref{fig:YIG}.

\section{Remarks on disorder effects}

In the main text,
we have remarked that 
details of the edge condition,
such as the presence or absence of disorder,
may change the boundary condition
for the wave function of magnons,
but the existence of the magnonic Casimir effect
remains unchanged.
Here,
we add a comment on disorder effects.
Since the magnonic Casimir energy
does not depend on the Bose-distribution function
[see Eqs.~\eqref{eqn:CasE}-\eqref{eqn:CasEcont}],
not only the low-energy magnon mode ($\sigma=+$)
but also its high-energy mode ($\sigma=-$)
in ferrimagnets
contributes to the magnonic Casimir effect.
Therefore, 
it can be expected that 
as long as disorder effects on the bulk
are weak enough that 
the high-energy mode is little influenced,
the existence of
the magnonic Casimir effect in ferrimagnets
remains unchanged.

\section{Thermal Casimir effect}

In the main text,
we have explained that 
thermal contributions to the Helmholtz free energy arise
at nonzero temperature.
Here, we add a remark on it.
Although its thermal contribution 
is called the ``thermal Casimir energy'',
there is a crucial distinction between the Casimir effect
and the thermal Casimir effect:
The zero-point energy,
which is the key concept of quantum mechanics
and plays a crucial role in the Casimir effect,
is absent in the thermal Casimir effect.
It should be noted that the zero-point energy is one of the most striking phenomenon of quantum mechanics in the sense that 
there are no classical analogs.
The Casimir effect arises from the zero-point energy
due to quantum fluctuations
and is not affected by temperatures,
whereas
the thermal Casimir effect arises from thermal fluctuations
and is exponentially suppressed at low temperatures.
The thermal Casimir effect vanishes at zero temperature,
whereas
the Casimir effect does exist even at zero temperature.
Thus, there is a significant distinction 
between the Casimir effect and the thermal Casimir effect.

\end{document}